\documentclass{iaus}
\usepackage{graphicx}

\title[Turbulence in Molecular ISM] 
{Turbulence in the Molecular Interstellar Medium}

\author[Heyer \& Brunt]   
{Mark H. Heyer$^1$ \& Chris Brunt$^2$}

\affiliation{$^1$Department of Astronomy, University of Massachusetts,
Amherst, MA 01003, USA \break
$^2$School of Physics, University of Exeter, Stocker Road,
EX4 4QL, United Kingdom\break
email: heyer@astro.umass.edu,brunt@astro.ex.ac.uk}

\pubyear{2006}
\volume{237}  
\pagerange{1}
\date{?? and in revised form ??}
\jname{Triggered Star Formation in a Turbulent ISM}
\editors{B. G. Elmegreen \& J. Palous, eds.}
\begin{document}

\maketitle

\begin{abstract}
The observational record of turbulence within the molecular 
gas phase of the interstellar medium is summarized.  We briefly 
review the analysis 
methods used to recover the 
velocity structure function from spectroscopic imaging and the
application of these tools on sets of cloud data.
These studies identify a near-invariant 
velocity structure function that is independent of local the environment 
and star formation activity.  
Such universality accounts for the 
cloud-to-cloud scaling law between the global line-width and 
size of molecular clouds found by Larson (1981) and 
constrains the 
degree to which supersonic turbulence can regulate star formation. 
In addition, the evidence for large scale driving sources necessary 
to sustain supersonic flows is summarized. 
\keywords{interstellar turbulence, molecular ISM, star formation}
\end{abstract}

\firstsection 
\section{Introduction}

Turbulent motions are commonly observed within several phases of the interstellar medium 
(see Elmegreen \& Scalo 2005). Within the molecular gas phase, turbulent gas flows 
are supersonic and possibly, super-Alfvenic, and play a dual role in 
the dynamics and evolution of these regions.  Turbulence can provide 
a non-thermal, macroscopic 
pressure that lends support against self-gravity.  In addition, compressible,
supersonic flows may promote star formation by generating density 
perturbations within the shocks of colliding gas streams that eventually 
evolve into self-gravitating or collapsing protostellar cores
(Padoan \& Nordlund 2002; Mac Low \& Klessen 2004).  

Spectroscopy of molecular line emission, especially the rotational lines 
of CO, have long been the primary measurement from which turbulence is 
defined.  In fact, supersonic motions are inferred from the
very first CO spectrum observed by Wilson, Jefferts, \& Penzias (1970)
in which there is a 5 km/s wide line core in addition to the 
broad 100 km/s wing component that was later attributed to a luminous
protostellar outflow.  The 5 km/s core is significant broader than the 
sound speed of molecular hydrogen assuming a temperature of 30 K.

\begin{figure}[hbt]
\begin{center}
\includegraphics[width=0.95\hsize]{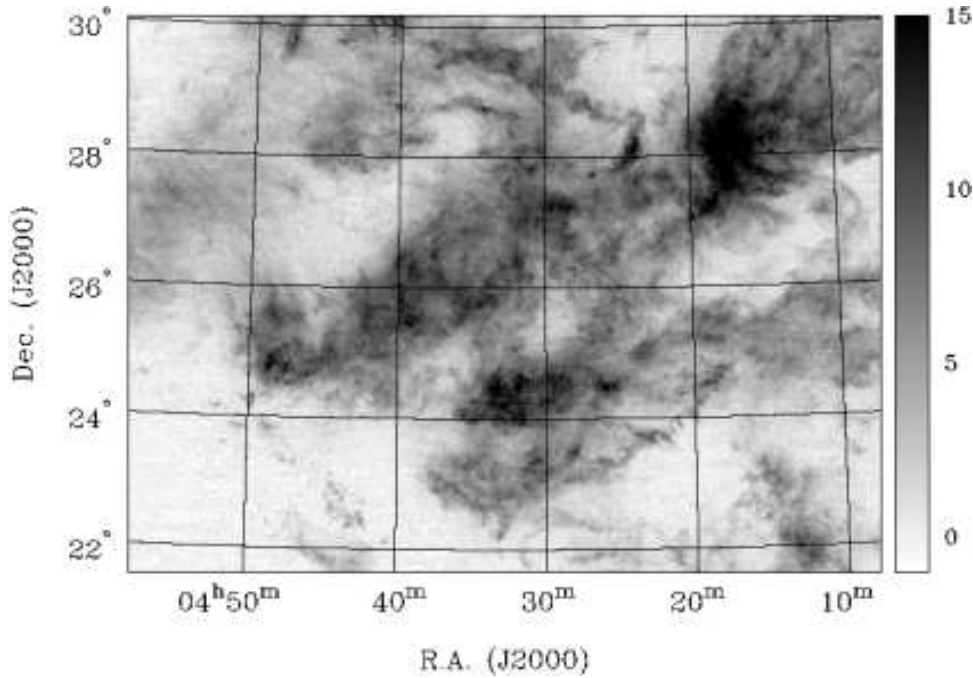}
\caption{An image of $^{12}$CO J=1-0 integrated emission from the Taurus 
Molecular Cloud observed with the {\it Five College Radio Astronomy 
Observatory} 14m telescope and {\it SEQUOIA} focal plane array.  The 
high spatial dynamic range reveals varying textures
across the cloud and clues to the prevailing physical processes.  
}
\label{taurus}
\end{center}
\end{figure}

Owing to advancing instrumentation at millimeter and submillimeter wavelengths, 
our ability to measure the distribution and 
kinematics of the molecular gas phase of the interstellar medium has greatly expanded
since that initial CO spectrum.
Sensitive, millimeter wave interferometers routinely 
probe the circumstellar environments about 
young stellar objects. Sensitive bolometer imaging arrays identify the sites
of protostellar and pre-protostellar cores (see Andre in these proceedings).  
Heterodyne focal plane arrays 
on single dish telescope enable the construction of high spatial dynamic
range imaging of molecular line emission (Heyer 1999).  
An example of such imaging is 
displayed in Figure~\ref{taurus}.  It reveals the varying texture of CO line 
emission imprinted by the effects of gravity, turbulence, 
and magnetic fields.  A diffuse, low surface brightness component 
extends across the field and contains localized ``streaks'' of emission 
that are
aligned along the local magnetic field direction.  The sequence of channel 
images show low column density 
material moving toward the dense, highly 
structured filaments that are more apparent in $^{13}$CO images and 
extinction maps. 
The challenge to the astronomer is to synthesize the 
information that is 
resident within these data cubes with suitable 
analysis tools to place these into a physical context in order to 
test and constrain model descriptions of 
turbulence within the molecular interstellar medium. 

\section{Velocity Structure Function}\label{sec:velstructfunction}
A primary goal in the study of ISM physics is to determine 
the degree of spatial correlation of velocities from observational data.
The velocity structure function, $S_q(\tau$),
defined as 
$$ S_q(\tau) = <|v(r)-v(r+\tau)|^q> $$
provides a statistical measure of the q$^{th}$ order of velocity differences 
of a field as a function of spatial displacement or lag, $\tau$. 
For q=2, $S_2(\tau)$, the autocorrelation function, $C(\tau)$, 
and the power spectrum 
are equivalent statistical measures of the velocity field. 
$S_2(\tau)$ is related to the autocorrelation function as 
$$ S_2(\tau) = 2(C(0)-C(\tau)) $$
and $C(\tau)$ is the Fourier transform of the 
power spectrum.   

Within the inertial range of 
a gas flow, the structure function is expected to vary as a power law with spatial lag,
$$ S_q(\tau) = {\delta}v^q(\tau) \propto \tau^{\zeta_q}.  $$
Taking the q$^{th}$ root of the structure function, this expression can 
be recast into an equivalent linear form, 
$ (S_q(\tau))^{1/q}=<{\delta}v>_q = v_\circ \tau^{\gamma_q} $ where $\gamma_q = \zeta_q/q$.
The power law index, $\gamma_q$, measures the degree of spatial correlation
and is predicted by model descriptions of turbulence (ex. $\gamma_3$=1/3 for 
Kolmogorov flow).  The normalization, $v_\circ$, is the amplitude 
of velocity fluctuations at a fixed scale and offers a convenient 
measure of the energy density of a flow.

While the expression for the structure function of a velocity field appears straightforward, 
the construction of $S(\tau)$ from observational data is, in fact,
quite challenging. 
Observers do not measure velocity {\it fields}, $v(r)$.   Rather, the basic 
unit of data is a spectrum of line emission that represents
a convolution and line of sight integration of density, velocity, and 
temperature. Furthermore, the effects of chemistry, opacity, and noise 
can mask or hide contributions to the line profile 
from features along the line of sight. 
Despite these limitations, there have been several 
demonstrated methods to recover the spatial statistics of GMC velocity 
fields from spectroscopic 
imaging data. \\
{\bf Analysis of Velocity Centroids}: A spectroscopic data cube can be  condensed into a 
2 dimensional image of centroid velocities determined from the set of line profiles. 
The spatial statistics of velocity centroids can be formally related to those of 
the 3 dimensional velocity field (see Ossenkopf etal 2006).   With the 
centroid velocity image, 
one can assess the power spectrum and hence structure function directly 
or apply a kernel to calculate the variance of centroid velocities 
over varying scales (Ossenkopf etal 2006).  This method works best 
under uniform density conditions (Brunt \& Mac Low 2004) or with 
an iterative scheme to account for density fluctuations 
within the measured power of the observed signal.\\
{\bf Velocity Channel Analysis}: 
Lazarian \& Pogosyan (2000) demonstrate a relationship between the power spectra
of measured line emission and the respective spectra of the density and 
velocity fields.  The relationship depends on the width of the 
velocity interval. By 
calculating the power spectra for both thick and thin velocity windows, one
can estimate the power law indices for both the density and velocity fields. \\
{\bf Principal Component Analysis}: The spectroscopic 
data cube is re-ordered onto a set of eigenvectors and eigenimages
(Heyer \& Schloerb 1997; Brunt \& Heyer 2002).  The 
eigenvectors describe the velocity differences in line profiles and the 
eigenimages convey where those differences occur on the sky.  The structure 
function is constructed from the velocity and angular scales determined 
from the 
set of respective eigenvectors and eigenimages that are significant 
with respect to the noise of the data.  To date, the results from PCA 
have been empirically linked to the velocity structure function 
parameters based 
on models under a broad range of physical and observational conditions
(Brunt \& Heyer 2002; Brunt etal 2003). 

\subsection{Universality of Turbulence}\label{sec:universality}

The three methods described in the previous section provide valuable tools to 
determine the velocity structure function for a singular interstellar cloud
from a set of spectroscopic imaging data.  However, for most 
observations, the statistical and systematic errors for the 
derived power law index for a given cloud is 
large ($\sigma_\gamma/\gamma \sim$ 10-20\%) and preclude a designation
of a turbulent flow type.
Moreover, given the broad diversity of environments and physical conditions
within the molecular ISM, any single measurement of a cloud is unlikely to 
characterize the complete population.
Therefore, it is imperative to analyze a large sample of 
molecular clouds to assess the impact of local effects and to 
identify trends and differences. 

\begin{figure}[hbt]
\begin{center}
\includegraphics[width=0.5\hsize]{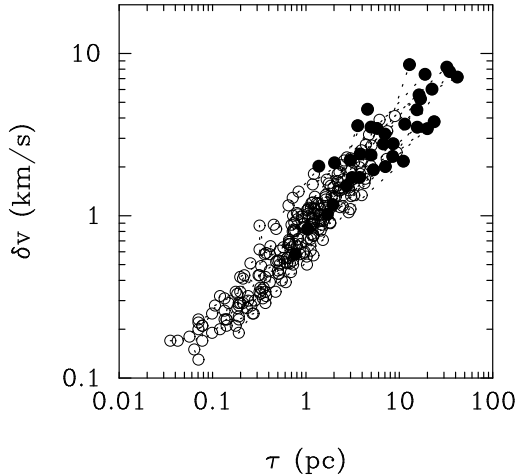}
\caption{The velocity structure functions for 29 clouds derived from 
PCA of $^{12}$CO J=1-0 data cubes (Heyer \& Brunt 2004).  
The nearly co-linear set of points 
attest to the near-invariant functional form of structure functions
despite the large range in size and star formation activity. The filled 
circles are the upper endpoints for each structure function and are equivalent
to the size and global line-width for each cloud. }
\label{uniplot}
\end{center}
\end{figure}

For Velocity Centroid Analysis, Miesch \& Bally (1994) analyzed a set of 
12 clouds or sub-regions within giant molecular clouds.  They determine a 
mean value of $\gamma$ to be 0.43$\pm$0.15.  Using PCA, 
Heyer \& Brunt (2004) studied 28 clouds in the Perseus and local 
spiral arms and found 
$\gamma$=0.49$\pm$0.15.  This mean 
value for $\gamma$ is consistent with highly supersonic turbulence 
in which the velocity field is characterized by ubiquitous 
shocks from converging gas streams.   The observed distribution 
of $\gamma$ would exclude a Kolmogorov description of 
incompressible turbulence unless the velocity fields are 
characterized by strong intermittency.  Moreover, they 
identified the surprising result that the scaling coefficient, $v_\circ$,
exhibits little variation from cloud to cloud, despite the large 
range in cloud sizes and star formation activity.  Effectively, 
when the individual structure functions are overlayed onto a single 
plot, they form a nearly co-linear set of points 
(see Figure~\ref{uniplot}).   

A necessary consequence of this 
universality is the Larson (1981) cloud-to-cloud size-linewidth 
relationship.  Basically, the upper endpoint of each individual 
structure function corresponds to the global size and line width 
of a cloud (filled points in Figure~\ref{uniplot}).  This set of 
endpoints are correlated only by the fact that the individual 
velocity structure functions are described by similar values for 
$\gamma$ and $v_\circ$.  If there were significant variations of
these parameters, then the scatter of points on the cloud-to-cloud
relationship would be much larger than is observed.  Using Monte Carlo 
simulations to model the scatter of line-width and size for 
GMCs in the inner Galaxy, Heyer \& Brunt (2004)
constrain the variation of $\gamma$ and $v_\circ$ to be less than 10-15\%. 

\section{Turbulent Driving Scales}\label{sec:turbdrive}
The measurements of velocity structure functions in the 
molecular ISM point to supersonic turbulent flows in which 
energy is dissipated in shocks.  Unless this energy is replenished 
within a crossing time,
the velocity field would 
evolve into a Kolmogorov flow comprised exclusively of solenoidal or 
eddy-like motions. The fact that we 
observe supersonic turbulence demonstrates that such driving sources must be 
present in the molecular ISM. 
Miesch \& Bally (1994) summarize candidate 
sources of energy that could 
sustain the observed turbulent motions.  These include sources that may be 
resident within the molecular cloud such as protostellar outflows  and 
intermediate and external sources such as HII regions, supernova remnants,
and Galactic shear.  
While all such sources make 
some contribution, it is important to assess whether any one 
process is the dominant source.  

As first noted by Larson (1981), 
the universality of velocity structure functions imply a common, 
{\it external} source of energy.  Otherwise, those regions with 
significant localized sources would exhibit significant departures from 
the observed universal relationship.  
However, GMCs with rich young clusters,
and OB stars show the same amplitude of velocity fluctuations as
low mass star forming clouds or even those few clouds with negligible 
star formation activity (Heyer, Williams, \& Brunt 2006). 
Either there are self-regulating processes independent of energy input 
scales that maintain this amplitude 
for most interstellar clouds or such internal energy sources contribute 
only a small fraction of the energy budget of a molecular cloud.

\begin{figure}[hbt]
\begin{center}
\includegraphics[width=0.9\hsize]{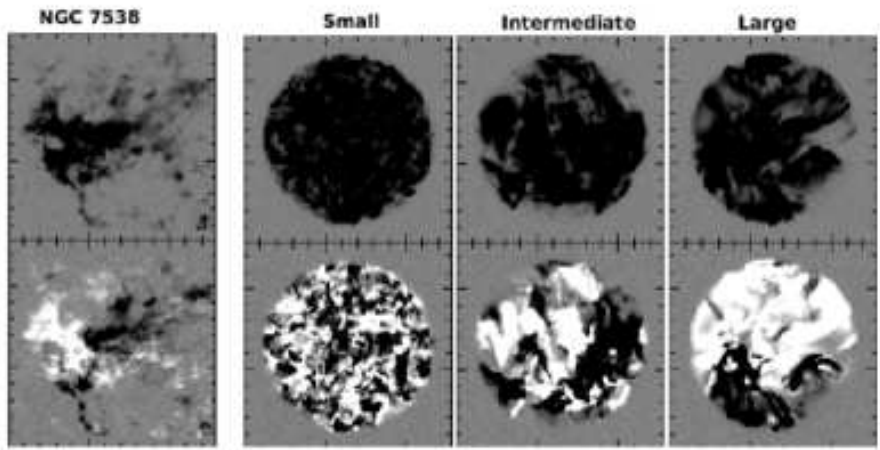}
\caption{The first two eigenimages derived from $^{12}$CO J=1-0 
emission from the NGC~7538 molecular cloud and model spectra from
computation simulations driven at small, intermediate, and 
large scales (Mac Low 1999).  Most observations of molecular clouds 
exhibit a dipole distribution in the second eigenimage that 
places most of the kinetic energy within the largest scales of a
cloud.  The observations are more congruent with computational 
simulations driven at scales larger than or equal to the size
of the cloud.}
\label{drive}
\end{center}
\end{figure}

Large scale driving is also implied by the observation that 
most of the kinetic energy of a cloud is distributed over the 
largest scales (Brunt 2003).  This is illustrated in Figure~\ref{drive},
which displays the first and second 
PCA eigenimages derived from $^{12}$CO J=1-0 
emission from the NGC~7538 molecular cloud.  The first eigenimage is 
similar to an integrated intensity image over the full velocity range 
of the cloud.  The second eigenimage exhibits a dipole-like distribution 
that identifies the large scale shear 
across the cloud.  All clouds studied by Heyer \& Brunt (2004) exhibit 
this dipole distribution in the second eigenimage. 
For comparison, we show the first two eigenimages calculated 
from simulated observations of velocity and density fields produced by 
computational models that are driven at small, intermediate, and large 
scales.  Using the ratio of characteristic scales determined from 
each eigenimage, one can quantitatively show that the observations 
are best described by a large scale driving force (Brunt 2003).
Protostellar outflows can have a significant but localized impact on 
a sub-volume of a cloud and can redistributed energy and 
momentum to large scales (see Bally in these proceedings).  However, 
it seems quite unlikely that an ensemble of widely distributed outflows 
within the cloud's volume, can generate 
the large scale shear that is observed within all molecular 
clouds analyzed to date.  

\section{Conclusions}\label{sec:concl}

Our understanding of turbulence in the molecular interstellar 
medium has greatly advanced over the last 10 years owing to 
more sophisticated computational simulations and ever improving 
observations.  However, there are more critical questions to 
address to improve these descriptions of turbulence and the 
role it plays in the star formation process.
\begin{itemize}
\item Does the shape of the velocity structure function for a given region 
change at spatial scales 
smaller than current resolution limits?
\item Does the universality of velocity structure functions extend to 
the extreme environment of the Galactic Center?
\item Are velocity fields of interstellar clouds anisotropic as 
predicted by the 
theory of strong, MHD turbulence (Goldreich \& Sridhar 1995)?
\end{itemize}

New telescopes, instrumentation, and analysis methods
will be required to address these questions.  ALMA will provide 
both sensitivity and angular resolution to investigate velocity 
structure functions at the smallest scales and for distant 
GMCs.  The Large Millimeter Telescope will offer the capability to 
study the low surface brightness component of the molecular ISM
to trace the transition from turbulent diffuse material 
to the dense proto-stellar and proto-cluster cores. 
These instruments, and others, offer exciting, scientific opportunities
to advance our knowledge of interstellar turbulence.

\begin{acknowledgments}
This work was supported by NSF grant AST 05-40852  to the Five College
Radio Astronomy Observatory. C.B holds a RCUK Academic Fellowship at the 
University of Exeter.  M.H. acknowledges support from the AAS
International Travel Grant Program.
\end{acknowledgments}

\end{document}